# New concepts for calibrating non-common path aberrations in adaptive optics and coronagraph systems

François Hénault
Institut de Planétologie et d'Astrophysique de Grenoble
Université Grenoble-Alpes, Centre National de la Recherche Scientifique
B.P. 53, 38041 Grenoble – France

## ABSTRACT

Non Common Path Aberrations (NCPA) are often considered as a critical issue in Adaptive Optics (AO) systems, since they introduce bias errors between real wavefronts propagating to the science detectors and those measured by the Wavefront Sensor (WFS). This is especially true when the AO system is coupled to a coronagraph instrument intended for the discovery and characterization of extra-solar planets, because useful planet signals could be mistaken with residual speckles generated by NCPA. Therefore, compensating for those errors is of prime importance and is already the scope of a few theoretical studies and experimental validations on-sky. This communication presents the conceptual optical design of an interferometric arrangement suitable to accurate NCPA calibration, based on two WFS cooperating in real-time. The concept is applicable to both classical imaging and spectroscopy assisted by AO, and to high-contrast coronagraphs searching for habitable extra-solar planets. Practical aspects are discussed, such as the choice of WFS and coronagraph types, or specific requirements on additional hardware components, e.g. dichroic beamsplitters.

**Keywords:** Adaptive optics, Wavefront sensor, Non common path aberration, Coronagraph, High contrast

## 1 INTRODUCTION

Non Common Path Aberrations (NCPA) are often considered as a critical issue in Adaptive Optics (AO) systems, since they introduce bias errors between real wavefronts propagating to the science detectors and those measured by the Wavefront Sensor (WFS). This is especially true when the AO system is coupled to a coronagraph instrument intended for the discovery and characterization of extra-solar planets, because useful planet signals could be mistaken with residual speckles generated by NCPA. The origin of those NCPA errors is schematically illustrated in Figure 1, showing the three essential components of an AO system, namely:

- A beam splitting plate (usually dichroics) picking part of the input optical beam and redirecting it towards the WFS,
- The WFS itself, providing an estimate of the input Wavefront Error (WFE) map noted $W(P)$[1],
- A deformable Mirror (DM) usable for adding specified distortions to the input WFE.

The amount of NCPA inside the system can be evaluated quantitatively by use of the following quantities:

$W_0(P)$     the input WFE on the DM, resulting from the addition of atmospheric disturbance (seeing) and optical aberrations of the telescope and other fore-optics,
$W_M(P)$     optical aberrations added by the DM at rest and relay optics, up to the dichroics beam splitter,
$W_1(P)$     the total WFE impinging on the beam splitter, equal to $W_0(P) + W_M(P)$,
$W_{R1}(P)$     the WFE reflected off the beam splitter towards the WFS (including eventual additional relay optics),
$W_2(P)$     additional aberrations on the way from beam splitter to the science detector.

---

[1] Most of WFS actually measure the partial derivatives of the WFE instead of the WFE itself, but this has no consequence on the validity of the principle presented here.

When the DM is at rest as sketched in Figure 1-a the WFE impinging the science detector is thus equal to $W_1(P) + W_2(P)$, while it is measured as $W_{M1}(P) = W_1(P) + W_{R1}(P)$ by the WFS. When operating in close loop (Figure 1-b) the deformations added by the DM to the input WFE may ideally be modelled as $-W_1(P) - W_{R1}(P)$[1]. It follows that an apparently corrected flat WFE is now measured by the WFS, while residual aberration still affect the science detector. Hence the basic NCPA expression is:

$$NCPA(P) = W_2(P) - W_{R1}(P). \qquad (1)$$

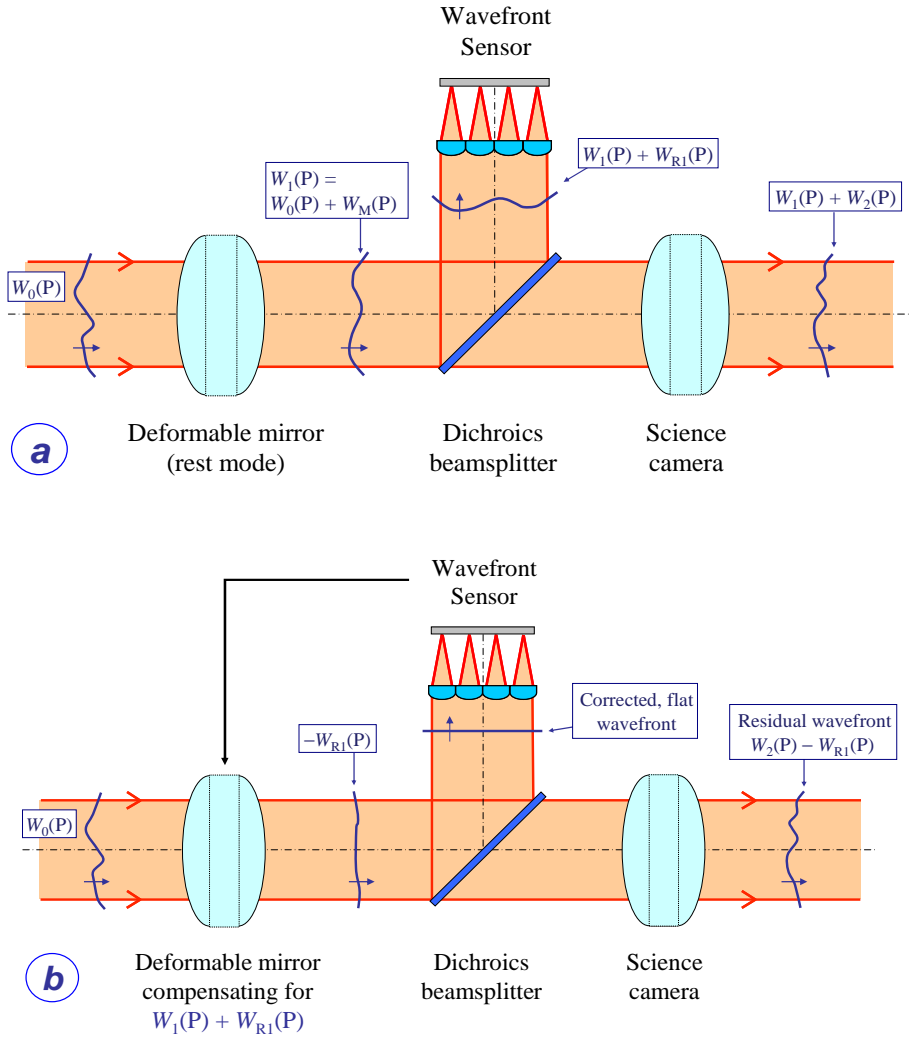

Figure 1: Illustrating NCPA errors inside an AO system. (a) Deformable mirror in rest mode. (b) Closed-loop mode.

Compensating for NCPA errors is of prime importance for high contrast instruments and has been the scope of a few theoretical and experimental studies. Basically two different types of solutions have been explored:

- Hardware modifications of the AO system. It may imply the development of new types of WFS [1] or modifications of the optical architecture of the instrument. Here the goal is to push the beam splitter closer to the science detector [2-3], and eventually behind a coronagraph phase mask [4]. Some of these concepts have already been partly validated on-sky [5].

---

[1] Neglecting important effects such as detection noise or DM responses, which are not the main scope of this paper.

- *A posteriori* correction of the NCPA and images acquired by the science detector [6-8]. They are out of the scope of the present study that only focuses at hardware solutions.

In this communication is presented the conceptual optical design of an interferometric arrangement suitable to accurate NCPA calibration, based on two WFS cooperating in real-time. The concept is presented in section 2 as well as applications to both classical imaging and spectroscopic instruments assisted by AO, or to high-contrast coronagraphs. Practical aspects are discussed in section 3, such as specific requirements on additional hardware component. The choice of the WFS and coronagraph types is also considered. Concluding remarks are given in final section 4.

## 2   CONCEPTUAL OPTICAL DESIGN

In this section is firstly presented the general concept for NCPA calibration (§ 2.1). Applications to spectroscopic instruments and coronagraphs are briefly described in the following subsections 2.2 and 2.3 respectively.

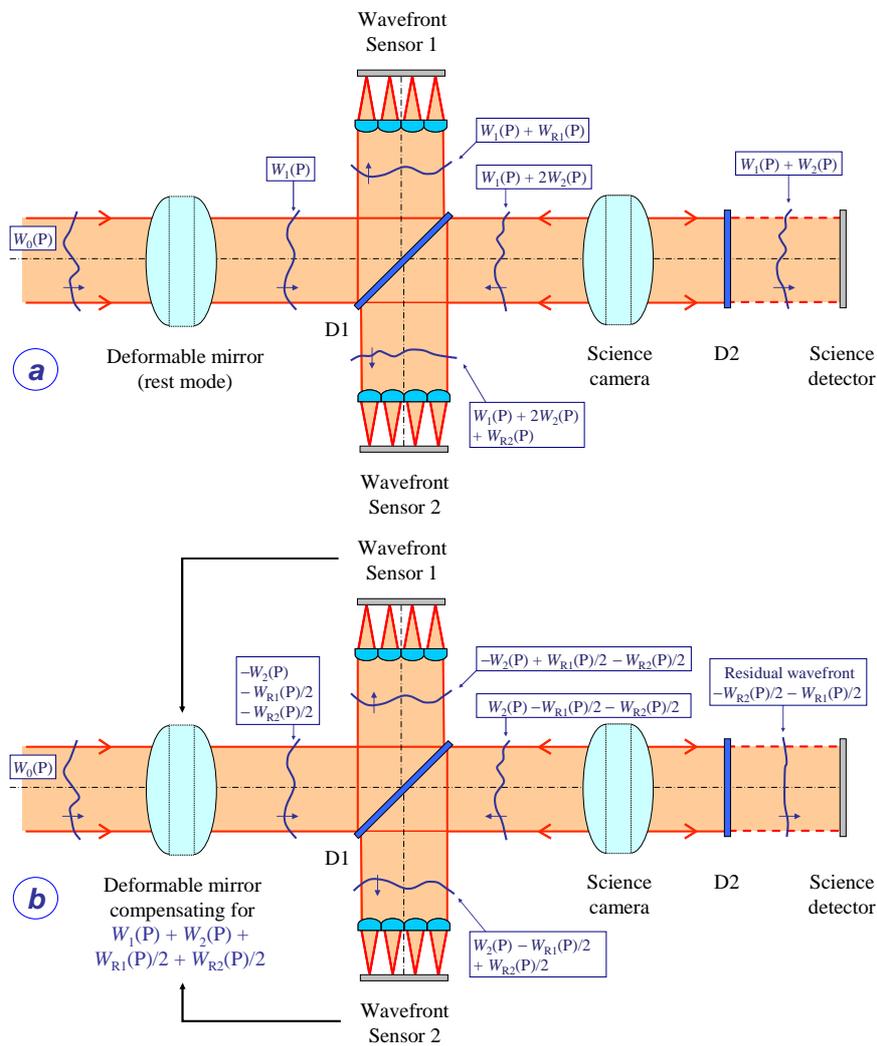

Figure 2: NCPA error reduction concept. (a) Deformable mirror in rest mode. (b) Closed-loop mode.

## 2.1 Principle

The proposed concept for NCPA calibration is illustrated in Figure 2, where the analytical expressions of the wavefronts are indicated at different locations in the optical system. It consists in building an interferometric arrangement between the first beam splitter, here and in the whole paper denoted D1, and a second dichroics beam splitter denoted D2. The latter shall be located as close as possible to the science detector, ideally after science camera optics as shown in the Figure. The optical quality of D2 is required to be the same as for an interferometer caliber, so that it reflects the incident wavefront $W_1(P) + W_2(P)$ through the camera optics without significant distortion. It is then reflected by dichroics beam splitter D1 towards a second wavefront sensor noted WFS-2, in the opposite direction to the first WFS (now and in the remainder of the text noted WFS-1). When the DM is at rest (see Figure 2-a) the WFE measured by WFS-2 is equal to $W_{M2}(P) = W_1(P) + 2W_2(P) + W_{R2}(P)$, where $W_{R2}(P)$ is the WFE reflected by D1 towards WFS-2. In the mean time WFS-1 should measure the same wavefront as in the previous section, i.e. $W_{M1}(P) = W_1(P) + W_{R1}(P)$. Here the basic idea consists in computing the arithmetical mean of both WFE measured simultaneously by WFS-1 and WFS-2, and using the result as an error signal $ES(P)$ for closed-loop operation:

$$ES(P) = \frac{W_{M1}(P) + W_{M2}(P)}{2} = W_1(P) + W_2(P) + \frac{W_{R1}(P) + W_{R2}(P)}{2}. \tag{2}$$

Hence the residual NCPA at the science detector are reduced to:

$$NCPA(P) = -W_{R1}(P)/2 - W_{R2}(P)/2. \tag{3}$$

Comparing Eq. 3 to Eq. 1 shows that better NCPA compensation should be achieved, since differential aberrations are reduced by a factor of two along WFS-1 optical path. More importantly, wavefront errors $W_2(P)$ due to the science camera optics are now replaced with the term $-W_{R2}(P)/2$ standing for differential aberrations between the first dichroics beam splitter D1 and WFS-2, also divided by a factor of two. Deeper minimization of the NCPA will also result from the symmetric beamsplitter configuration, as explained in subsection 3.1.2.

## 2.2 Application to imaging or spectroscopic instruments

Possible applications of the concept to a spectroscopic instrument are depicted in Figure 3. Two variants are presented:

- Figure 3-a shows a conventional long-slit spectrograph located behind the dichroics beamsplitter D1 and both wavefront sensors WFS-1 and WFS-2. Here D1 is inserted into a converging optical beam, which is the most encountered case (though not necessarily recommended) and may also influence the choice of wavefront sensors type (see § 3.2). The spectrograph itself is represented in the right side of Figure 3. Without loss of generality, its dispersive element is depicted as a grism located between the collimating optics L2 and camera optics L3. The dichroics beamsplitter D2 is located at the spectrograph pupil, here the entrance face of the grism. This configuration allows calibrating all differential aberrations between the wavefront sensors WFS-1 and WFS-2 and the dispersive element. Therefore only the spectrograph camera optics should contribute to NCPA errors.

- Figure 3-b shows a variant where the beamsplitter D2 is located much closer to the science detector, i.e. after camera optics L3, where the beam is converging. It implies that D2 should be a meniscus made of two spherical surfaces, and that the curvature radius of the coated surface is equal to the focal length of the camera. In that way NCPA correction now includes the WFE introduced by the dispersive element and camera optics. From a practical point of view, it can be noted that wavefront sensing is performed in a reduced spectral channel than science observations. It follows that D2 should be located below the science detector (see Figure 3-b where the scientific spectral range is indicated as $\lambda_2$ to $\lambda_3$). Thus the D2 dichroics could be replaced with a simpler all-reflective coating. The whole component could also be integrated into the detector package as a tooling ball located just below the sensitive area (this last option is not shown in the Figure). However this optical configuration suffers from a major drawback, since only a small spectral bandwidth can be reflected back from D2 to WFS-2, which should result in significant losses in terms of signal-to-noise ratio.

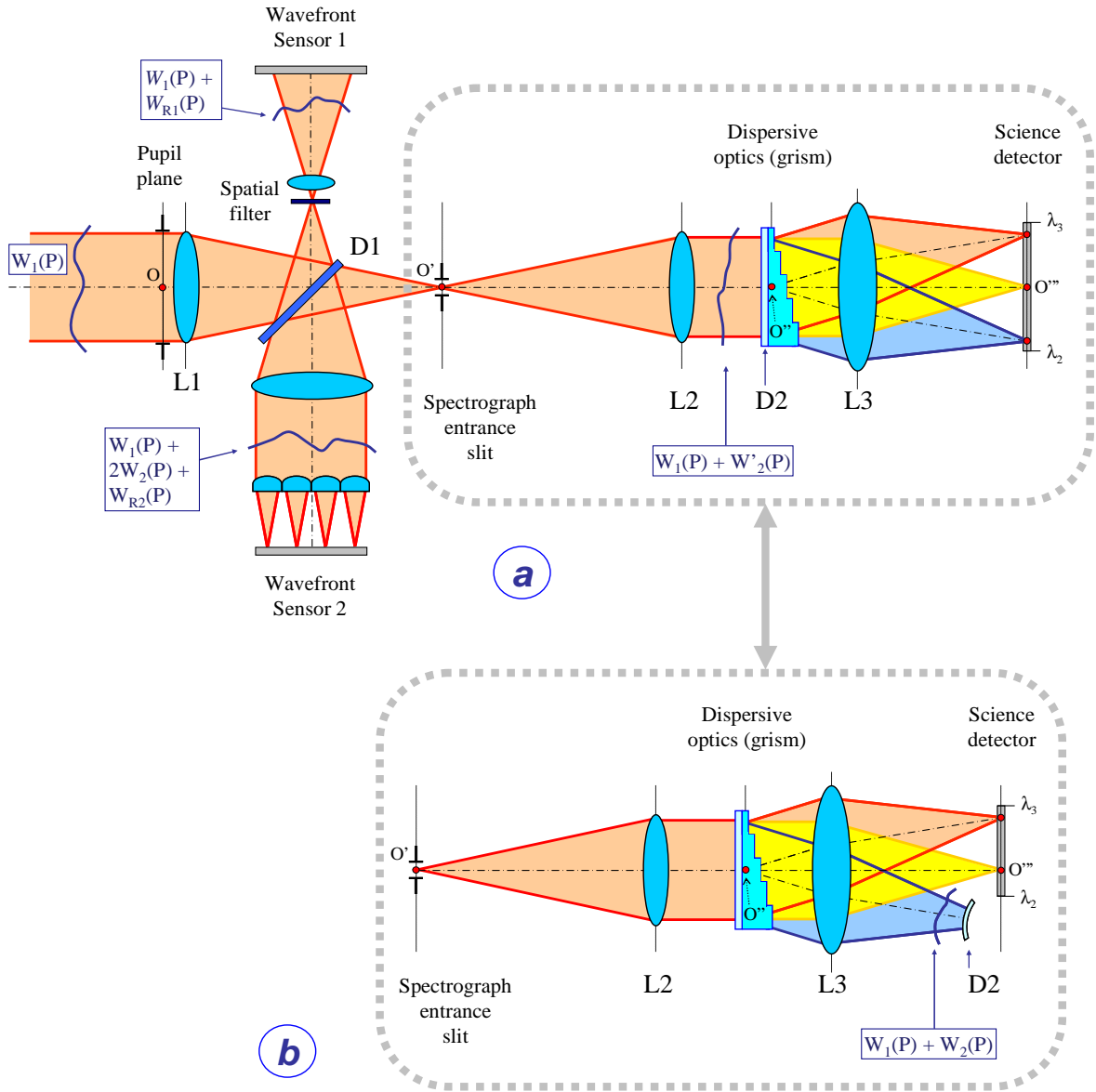

Figure 3: Application to a spectroscopic instrument with D2 dichroics located at the dispersive element (a) or near the science detector (b).

## 2.3 Application to high contrast instruments – Coronagraphs

Applications to a coronagraph instrument are depicted in Figure 4. Here again two variants are examined, depending on the type of the coronagraph.

### 2.3.1 Pupil apodization coronagraphs (Figure 4-a)

In a most general sense, pupil apodization coronagraphy consists in modifying the complex amplitude collected by the telescope by means of an optically diffractive component added into a pupil plane. Two sub-classes can be distinguished:

- The optical component is purely transmitting, i.e. it only affects the pupil transmission map. Transmission changes over the surface of the pupil can either be continuous [9] or binary [10], eventually leading to the definition of highly complex apodizing patterns [11].

- The optical component only introduces phase gradients into the pupil plane, possibly using a deformable mirror [12] or pre-determined phase plates [13].

In both cases the searched effect is to modify the Point-Spread Function (PSF) of the instrument, so that one or several dark areas (or "dark holes") are created around the central lobe of the PSF. Thus faint planets located off-axis from the central star should become detectable in those dark areas. In such type of coronagraph the central star is not removed from the final image and very few diffracted rays are propagated trough the optical system. It follows that the presence of a "Lyot" stop is not mandatory. Figure 4-a illustrates the application of the NCPA calibration method to this type of coronagraph. The arrangement of the cooperating wavefront sensors WFS-1 and WFS-2 and dichroics beamsplitter D1 is similar as described in section 2.2. Since WFE control is very critical for coronagraphic applications, it is again desirable to push beamsplitter D2 as close as possible to the science detector: as in § 2.2, it shall be a meniscus of curvature radius equal to the focal length of the science camera L3. Basic requirements for the coatings of dichroic plates D1 and D2 are given in § 3.1.1.

It may be noted that the same type of coronagraph has been selected for the High Contrast Module (HCM) that will be integrated into the first-light ELT instrument HARMONI [2-3]. The HCM includes a set of binary apodizing mask of high complexity, which may preclude the choice of certain types of wavefront sensors for WFS-2. Upgrading the current HCM design to incorporate the NCPA calibration concept presented in this paper could be an exciting study in the future.

### 2.3.2 *Phase mask coronagraphs (Figure 4-b)*

The main difference between this type of coronagraph and the previous one is that the diffractive optical component is set into an image plane of the optical system rather than in a pupil plane. In a vast majority of cases this diffractive element only modifies the phase of the incident complex amplitude, hence their generic name of Phase Mask Coronagraph (PMC). So far, the most well-known PMCs probably are the Roddier and Roddier, four-quadrants, and vortex coronagraphs, whose diffractive properties were discussed extensively in Refs. [4] and [14]. They show that most of the starlight is redirected outside of the pupil area (as schematized by the orange beam in Figure 4-b), which entails the presence of a Lyot stop preventing starlight from reaching the science detector[1]. This case is probably the most demanding in terms of optical design, because:

- Collimating optics L2 and camera optics L3 should have higher numerical apertures than required for the science beam, because necessary information for wavefront sensing is now spread over the whole diffracted beam (shown in orange color). Numerical simulations in Ref. [4] demonstrated that for efficient WFE reconstruction their apertures should be oversized by a factor of three at least. Such requirement may turn even more critical if a spectrograph has to be coupled with the coronagraph.

- It also implies the need for a "dichroics Lyot stop" $D_L$ with other potential manufacturing difficulties, since the beam splitting area should be restricted to an external ring blocking the science spectral range (above $\lambda_2$) and transmitting WFS-2 spectral band only (below $\lambda_2$).

Regardless of such practical issues, it seems that this configuration is well-suited to efficient NCPA compensation applied to phase mask coronagraphs. It must be noted however that the basic reasoning involves a strong theoretical hypothesis:

- Light is fundamentally considered as a complex amplitude wave having no definite sense of propagation. This means that complex amplitude distributions in different pupil and image planes only are Fourier or inverse Fourier transforms of each other, whatever their actual locations and disregarding the alternative model of photons being successively diffused forward and backward by the phase mask. Since that last model would probably conclude that WFE measurements are not feasible with WFS-2, a practical realization of such an experiment may contribute to bringing new insights on the nature of light.

---

[1] Same considerations are also applicable to the historical Lyot coronagraph [15], where a central occulting spot is used instead of phase masks.

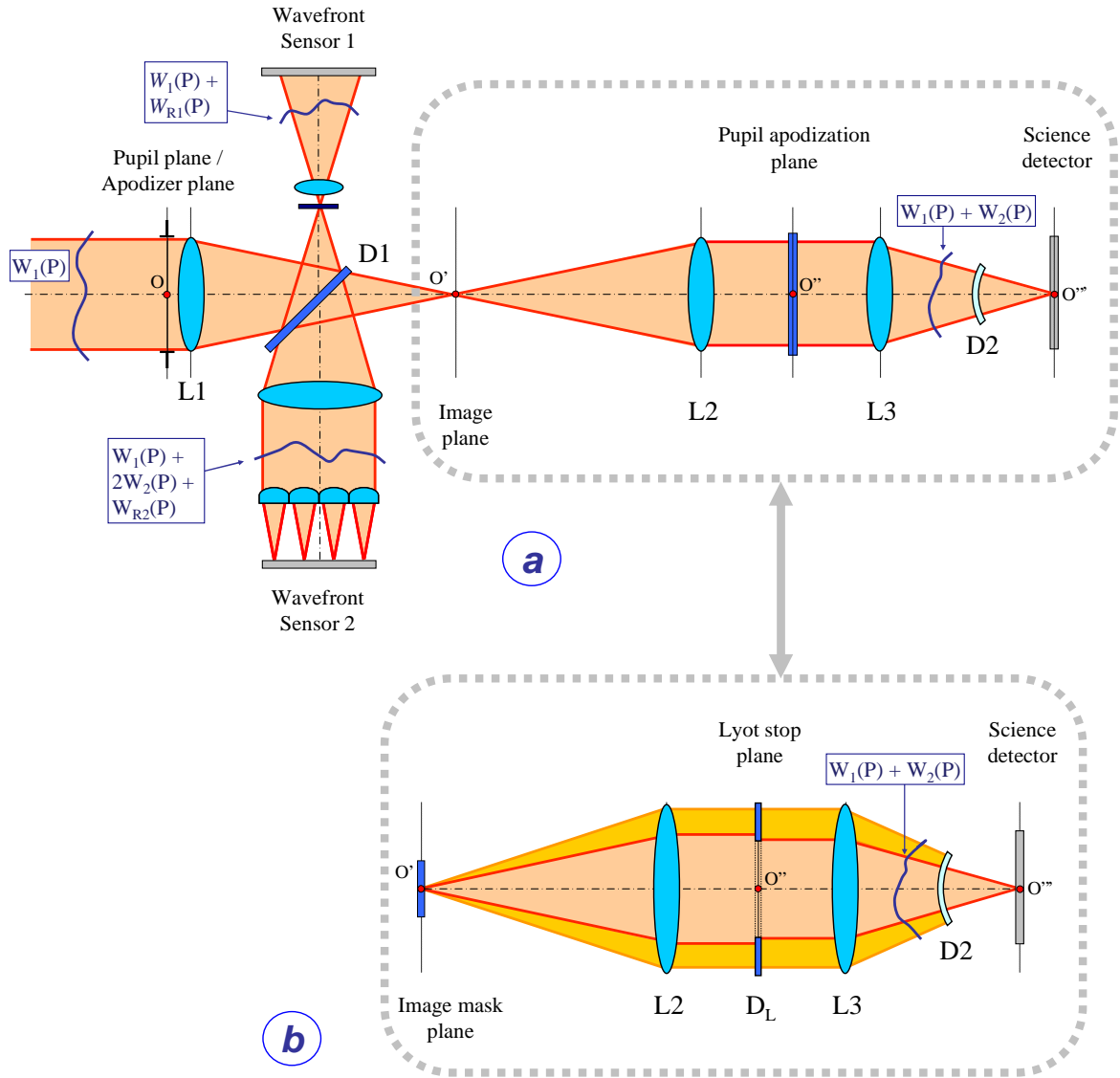

Figure 4: Applications to a coronagraph instrument of ppupil apodization type (a) or phase mask type (b).

## 3    DISCUSSION

In this section are firstly presented some preliminary requirements for the employed dichroic plates (§ 3.1), since they probably are the most critical components for ensuring the success of the NCPA calibration setup presented in this paper. Selecting the types of both wavefront sensors WFS-1 and WFS-2 is also briefly discussed in § 3.2.

## 3.1 Preliminary specifications for dichroic plates

Spectral transmission and image quality requirements of dichroics D1 and D2 are discussed in subsections 3.1.1 and 3.1.2 respectively.

### 3.1.1 Spectral transmission

In Figure 5 is presented a simplified radiometric budget for the NCPA calibration setup. Only dichroics D1 and D2 are considered, other optics being assumed ideal with a transmission equal to unity. The following scientific notations are employed:

| | |
|---|---|
| $[\lambda_1 - \lambda_2]$ | Operating spectral range of WFS-1 and WFS-2 |
| $[\lambda_2 - \lambda_3]$ | Operating spectral range of science beam |
| $T_1(\lambda)$ | Spectral transmission curve of dichroic beamsplitter D1 |
| $R_1(\lambda)$ | Spectral reflection curve of dichroic beamsplitter D1. Assuming D1 to be lossless $R_1(\lambda) = 1 - T_1(\lambda)$ |
| $T_2(\lambda)$ | Spectral transmission curve of dichroic beamsplitter D2 |
| $R_2(\lambda)$ | Spectral reflection curve of dichroic beamsplitter D2. Assuming D2 to be lossless $R_2(\lambda) = 1 - T_2(\lambda)$ |

It is also assumed that wavefront sensing operates at shorter wavelengths than the science beam, thus $\lambda_1 < \lambda_2 < \lambda_3$. In Figure 5-a is firstly shown a radiometric map indicating the effective transmission and reflection coefficients inside the system. Figure 5-b and 5-c illustrate the ideal spectral transmission and reflection curves of dichroics beamsplitter D1: its basic requirement is to transmit and reflect half of the incident beam in the $[\lambda_1 - \lambda_2]$ spectral range, i.e. $T_1(\lambda) = R_1(\lambda) = 0.5$. It also perfectly transmits higher wavelengths through the main optical system. The dichroic plate D2 has a more conventional coating with a cutoff wavelength equal to $\lambda_2$. Its spectral curves are sketched in Figure 5-d and 5-e, reflecting all wavelengths shorter than $\lambda_2$ back through the optical system, and letting higher wavelengths finally reach the science detector. Combining the four previous spectral curves allows determining the spectral radiometric characteristics of the beams finally impinging WFS-1, WFS-2 and the science detector. The results are summarized in Table 1, once again neglecting any other radiometric losses not originating from D1 and D2. It follows that in their common spectral range $[\lambda_1 - \lambda_2]$ WFS-1 could potentially collect one half of the available optical power, while WFS-2 is limited to one fourth of it.

Table 1: Achieved radiometric budgets for WFS-1, WFS-2 and science detector beams.

| Power collected by: | Analytic expression | Fraction of incident power | Spectral range | See Figure: |
|---|---|---|---|---|
| WFS-1 | $R_1(\lambda)$ | 0.5 | $[\lambda_1 - \lambda_2]$ | 5-b |
| WFS-2 | $R_1(\lambda) R_2(\lambda) T_1(\lambda)$ | 0.25 | $[\lambda_1 - \lambda_2]$ | 5-g |
| Science detector | $T_1(\lambda) T_2(\lambda)$ | ~ 0.96 [1] | $[\lambda_2 - \lambda_3]$ | 5-f |

Let us conclude this section with the two following remarks:

1) The previous discussion is fully applicable to the "basic" configurations of Figure 3-a and Figure 4-a, where the NCPA compensation method is implemented into the first stage of a spectrometer or to pupil apodization coronagraph. It is also applicable to the alternative spectrographic configuration of Figure 3-b for what concerns the dichroic plate D1 (D2 being replaced with a simple spherical mirror). Finally, the PMC configuration described in Figure 4-b involves an additional dichroics Lyot stop $D_L$ whose radiometric characteristics were briefly discussed in § 2.3.2.

2) Constraining WFS-1 to operating in the same spectral range as WFS-2 is not absolutely necessary. One may wish to extend its spectral band below $\lambda_1$ (as represented by dotted lines in Figure 5-b and 5-c) in order to maximize the collected optical power. Care should be taken however to limit chromatic NCPA that may result from the subsequent mismatch between WFS-1 and WFS-2 spectral ranges.

---

[1] Assuming $T_1(\lambda) = T_2(\lambda) = 0.98$, which seems more realistic than the unitary values sketched in Figure 5.

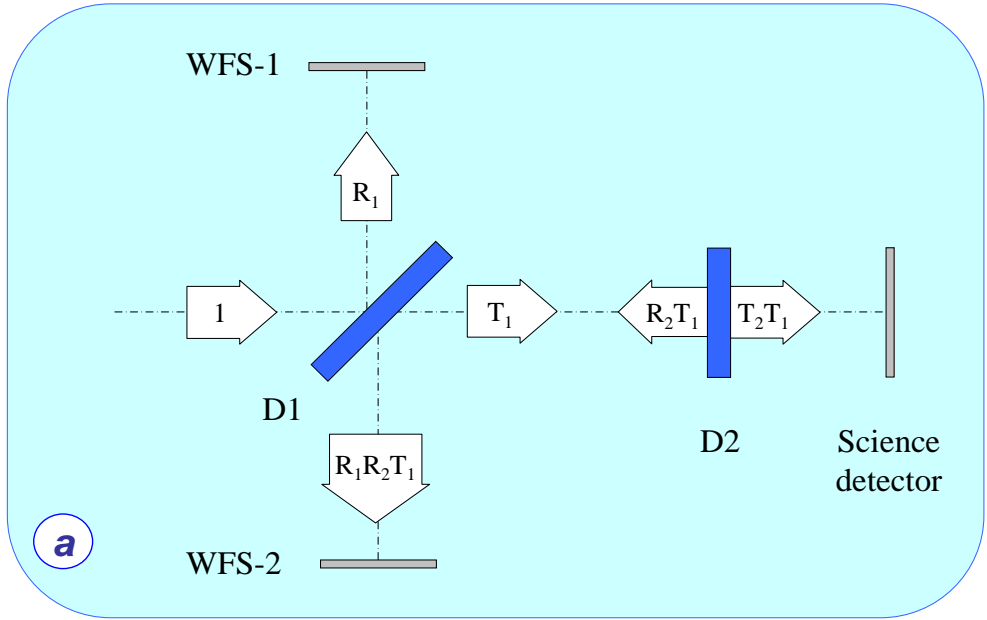

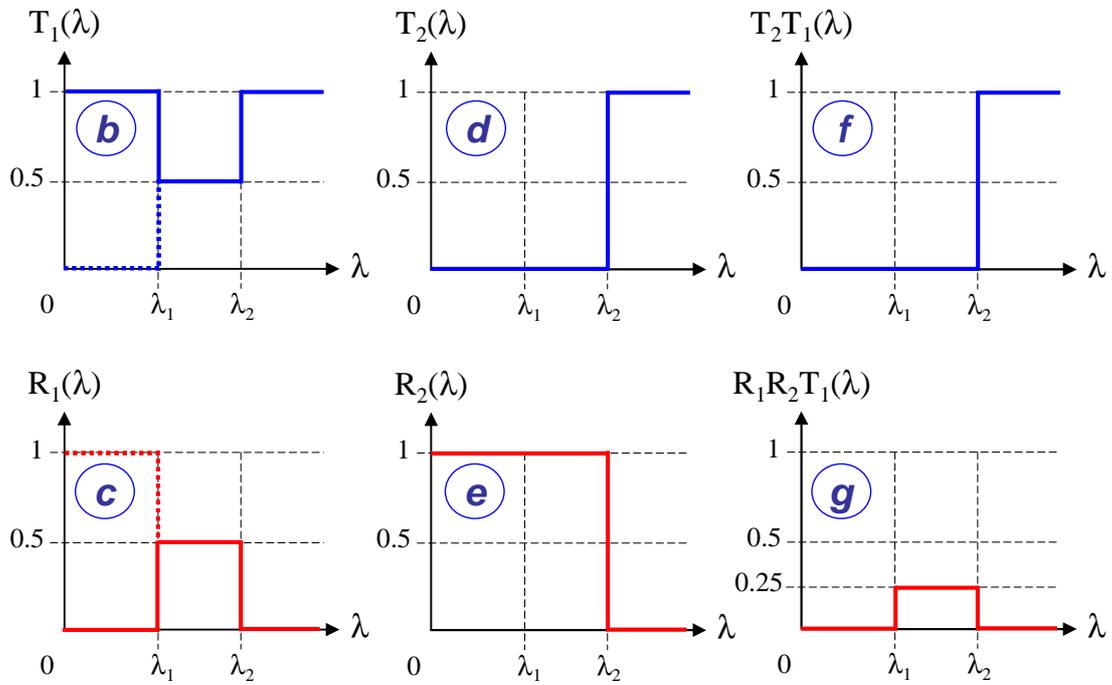

Figure 5: Radiometric budget and preliminary specifications of the NCPA calibration setup. (a) Transmission and reflection map of the system. (b) Spectral transmission curve of dichroic D1. (c) Spectral reflection curve of dichroic D1. (d) Spectral transmission curve of dichroic D2. (e) Spectral reflection curve of dichroic D2. (f) Global spectral transmission of the science beam. (g) Global spectral transmission of WFS-2 beam.

*3.1.2 Image quality*

Since the NCPA calibration method described in this paper is based on an interferometric arrangement, one may expect beamsplitters D1 and D2 to be subject to stringent optical quality requirements. This is especially true for D2: assuming a global NCPA calibration requirement of $\lambda/20$ RMS, its manufacturing and polishing accuracy should typically be $\lambda/40$

RMS, which is comparable to those of an interferometer caliber. However the case of beamsplitter D1 should be more favorable since its main function is to feed WFS-1 and WFS-2 simultaneously. In such arrangement, the choice of a symmetric dichroics beamsplitter seems to be the most natural and advantageous, but requires a specific study. Let us then consider the symmetric beamsplitter D1 depicted in Figure 6, and make use of the following notations:

$\Delta_1(P)$    Flatness error at the entrance surface of beamsplitter D1
$\Delta_R(P)$    Flatness error at the reflective dichroic surface of beamsplitter D1
$\Delta_2(P)$    Flatness error at the exit surface of beamsplitter D1

Neglecting the "cosine effect" due to its 45 deg. inclination, the wavefronts reflected by D1 towards WFS-1 and WFS-2 write respectively:

$$W_{R1}(P) \approx +2n(\lambda)\,\Delta_R(P) + 2(n(\lambda)-1)\,\Delta_1(P), \quad \text{and:} \tag{4a}$$

$$W_{R2}(P) \approx -2n(\lambda)\,\Delta_R(P) - 2(n(\lambda)-1)\,\Delta_2(P), \tag{4b}$$

with $n(\lambda)$ the refractive index of the dichroics plate. Then from Eq. 3 the residual NCPA should be:

$$NCPA(P) \approx (n(\lambda)-1)[\Delta_2(P) - \Delta_1(P)], \tag{5}$$

where the most important errors terms proportional to $\Delta_R(P)$ and originating from optical defects of the dichroic surface cancel each other. Residual NCPA are thus proportional to $\Delta_2(P) - \Delta_1(P)$, i.e. to the parallelism between the entrance and exit faces of the beamsplitter. Here the global NCPA requirement of $\lambda/20$ RMS should be translated into a parallelism specification around $\lambda/20$ RMS. Therefore the manufacturing requirements of dichroics D1 should be relaxed significantly with respect to those of dichroics D2.

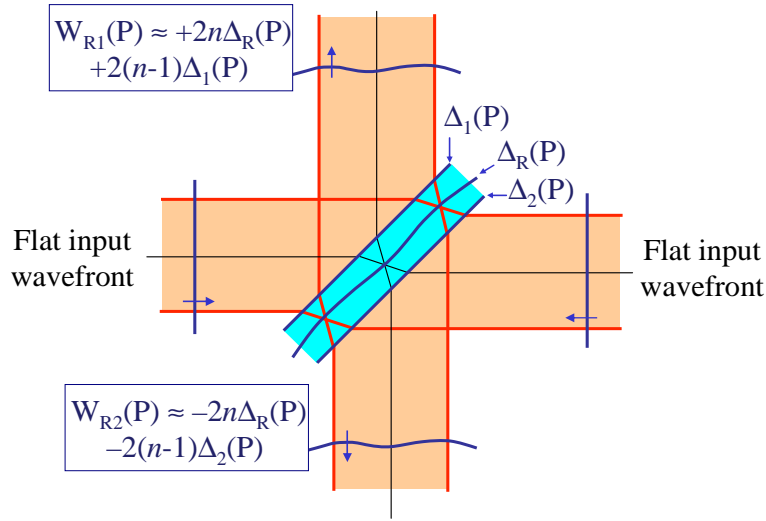

Figure 6: Symmetric beamsplitter configuration and its reflected wavefronts.

## 3.2   Choice of wavefront sensors

The optimal choice of a wavefront sensor in AO systems is a vast topic that has been the subject of extensive literature. Here it is only discussed in light of the proposed NCPA compensation method. Hence the main concern should be to restrict the number of additional optical components located between dichroics D1 and WFS-1, on the one hand, and between D2 and WFS-2, on the other hand. The choice of the WFS type also depends on the location of the dichroics beamsplitter in the optical train of the instrument, as schematically illustrated in Figure 7:

- If the dichroics is inserted into a parallel beam as sketched in Figure 7-a, both wavefronts reflected towards WFS-1 and WFS-2 are parallel, and the best choice seem to be the classical Shack-Hartmann WFS [16] that can be integrated into the reflected beam without additional optics.

- The use of an image plane WFS could also be envisaged, as shown in Figure 7-b where it is presented schematically as a spatial filter associated to focal plane optics. Those WFS could be of a few different types, such as the reverse Hartmann [17] or Zernike phase-mask WFS [1]. It implies however that focusing optics must be inserted before the WFS, therefore introducing additional NCPA.

- Conversely, if the dichroics is inserted into a converging or diverging beam, an image plane WFS could be integrated just after the dichroics D1 as shown in Figure 7-c. It is expected that the presence of focal plane optics will add a negligible amount of NCPA. Using the Shack-Hartmann WFS in that configuration would impose to add collimating optics (Figure 7-d), again with the risk of increasing NCPA.

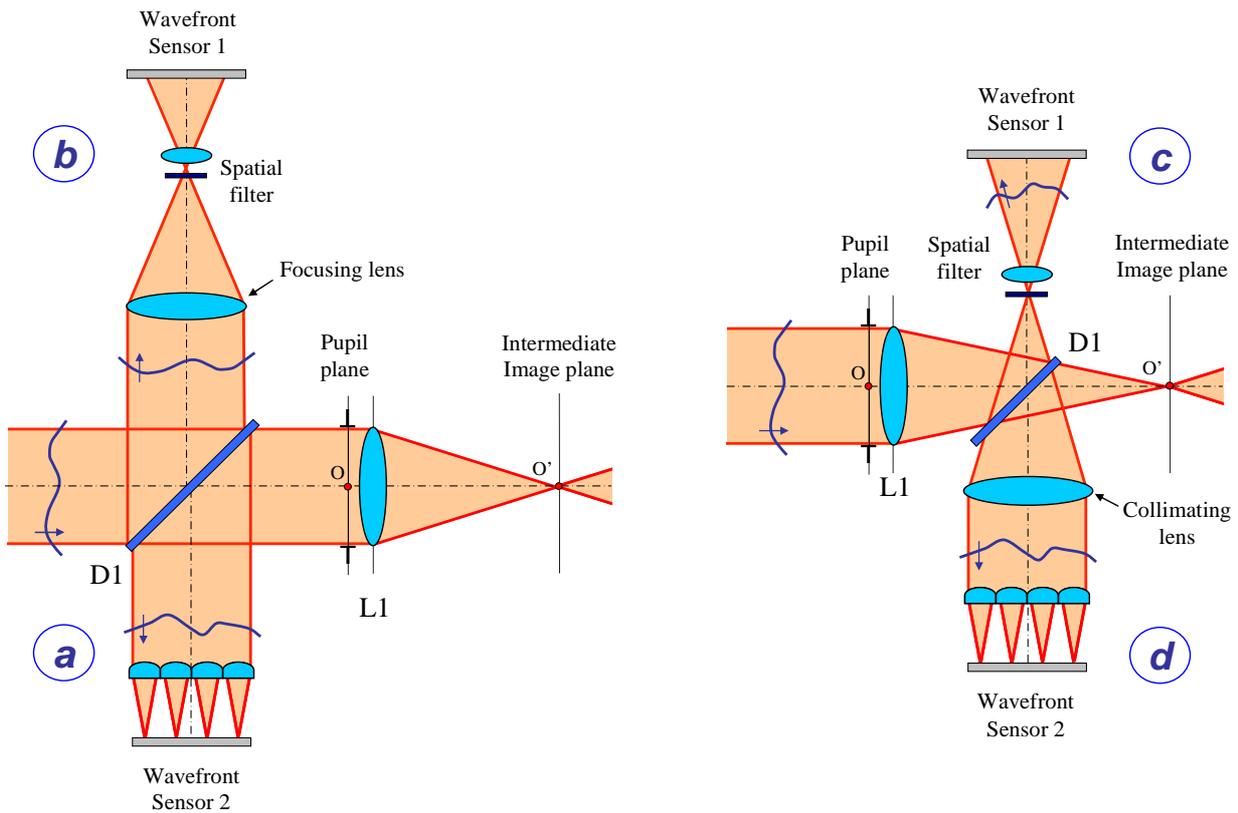

Figure 7: Possible choices and arrangements for wavefront sensors WFS-1 and WFS-2.

## 4 CONCLUSION

In this communication was explored a new method for calibrating NCPA inside AO systems. Starting from a classical setup, the method consists in building an interferometric arrangement that makes use of two custom-made dichroics beamsplitters and of two different WFS cooperating in real-time. Different configurations were described and discussed, applicable to both classical spectroscopy assisted by AO and to high-contrast coronagraphs searching for habitable extra-solar planets. Preliminary requirements were defined for the most critical optical components, clearly identified as the dichroics beamsplitters. One of them (D1) should have radiometric characteristics similar to those of an interferometric beamsplitter in the wavefront sensing spectral range, but presents some relaxed optical manufacturing requirements. Conversely, the second one (D2) has a standard dichroic coating, bur stringent image quality requirements comparable to those of an interferometer caliber.

## REFERENCES

[1] M. N'Diaye, K. Dohlen, T. Fusco, B. Paul, "Calibration of quasi-static aberrations in exoplanet direct-imaging instruments with a Zernike phase-mask sensor," Astronomy and Astrophysics vol. 555, n°A94 (2013).
[2] A. Carlotti, F. Hénault, K. Dohlen, J.F. Sauvage, P. Rabou, Y. Magnard, A. Vigan, D. Mouillet, P. Vola, G. Chauvin, M. Bonnefoy, T. Fusco, K. El Hadi, N. Thatte, F. Clarke, I. Bryson, H. Schnetler, M. Tecza, C. Vérinaud, "System analysis and expected performance of a high-contrast module for ELT-HARMONI," Proceedings of the SPIE vol. 10702 (2018).
[3] F. Hénault, A. Carlotti, P. Rabou, Y. Magnard, E. Sradler, D. Mouillet, G. Chauvin, M. Bonnefoy, J.F. Sauvage, K. Dohlen, A. Vigan, T. Fusco, K. El Hadi, F. Clarke, N. Thatte, I. Bryson, H. Schnetler, M. Tecza, C. Vérinaud, "Opto-mechanical design of a High Contrast Module (HCM) for HARMONI," Proceedings of the SPIE vol. 10702 (2018).
[4] F. Hénault, A. Carlotti, C. Vérinaud, "Phase-shifting coronagraph," Proceedings of the SPIE vol. 10400, n° 104001J (2017).
[5] A. Vigan, M. N'Diaye, K. Dohlen, J. Milli, Z. Wahhaj, J.-F. Sauvage, J.-L. Beuzit, R. Pourcelot, D. Mouillet, G. Zins, "On-sky compensation of noncommon path aberrations with the ZELDA wavefront sensor in VLT/SPHERE," Proc. SPIE vol. 10703, n°107035O (2018).
[6] J.-F. Sauvage, T. Fusco, G. Rousset, C. Petit1, "Calibration and precompensation of noncommon path aberrations for extreme adaptive optics," J. Opt. Soc. Am. A vol. 24, p. 2334-2346 (2007).
[7] D. Ren, B. Dong, Y. Zhu, D. J. Christian, "Correction of non-common-path error for extreme adaptive optics," Publications of the Astronomical Society of the Pacific vol. 124, p. 247-253 (2012).
[8] M. Lamb, D. R. Andersen, J.-P. Véran, C. Correia, G. Herriot *et al*, "Non-common path aberration corrections for current and future AO systems," Proc. SPIE vol. 9148, n° 914857 (2014).
[9] C. Aime, "Principle of an achromatic prolate apodized Lyot coronagraph;" Publications of the Astronomical Society of the Pacific vol. 117, p. 1012-1019 (2005).
[10] N.J. Kasdin, R.J. Vanderbei, M.G. Littman, D.N. Spergel, "Optimal one-dimensional apodizations and shaped pupils for planet finding coronagraphy;" Applied Optics vol. 44, p. 1117-1128 (2005).
[11] A. Carlotti, R. Vanderbei, N. J. Kasdin, "Optimal pupil apodizations of arbitrary apertures for high-contrast imaging;" Optics Express vol. 19, p. 26796- 26809 (2011).
[12] F. Malbet, J. W. Yu, M. Shao, "High-dynamic-range imaging using a deformable mirror for space coronography," Publications of the Astronomical Society of the Pacific vol. 107, p. 386-398 (1995).
[13] M. A. Kenworthy, J. L. Codona, P. M. Hinz, J. R. P. Angel, A. Heinze, S. Sivanandam, "First on-sky high-contrast imaging with an apodizing phase plate," Astrophysical Journal vol. 660, p. 762-769 (2007).
[14] F. Hénault, "Analysis of azimuthal phase mask coronagraphs," Optics Communications vol. 423, p. 186-199 (2018).
[15] B. Lyot, "The study of the solar corona and prominences without eclipses," Monthly Notices of the Royal Astronomical Society vol. 99, p. 580-594 (1939).
[16] R. V. Shack, B. C. Platt, "Production and use of a lenticular Hartmann screen," J. Opt. Soc. Am. vol. 61, p. 656 (1971).
[17] F. Hénault, "Fresnel diffraction analysis of Ronchi and reverse Hartmann tests," JOSA A vol. 35, p. 1717-1729 (2018).